\documentstyle[aps,preprint]{revtex}
\begin{document} 
\bigskip\bigskip
\centerline{\bf NATURE OF THE BACKGROUND ULTRAVIOLET 
RADIATION FIELD}
\centerline{\bf AT HIGH REDSHIFTS}
\bigskip \bigskip \bigskip 
\centerline{Archana Samantaray and Pushpa Khare}
\centerline{Physics Department, Utkal University} 
\centerline{Bhubaneswar, 751004, India}
\centerline{Running Title: Background UV Radiation field}
\newpage
\centerline{\bf ABSTRACT}

We have tried to determine the flux of the ultraviolet background radiation
field from the column density ratios of various ions in several absorption
systems observed in the spectra of QSOs.  We find that in most cases the flux is
considerably higher than what has been estimated to be contributed by the AGNs.
The excess flux could originate locally in hot stars.  In a few cases we have
been able to show that such galactic flux can only contribute a part of
the total required flux. The results suggest that the background gets a 
significant contribution from an unseen QSO population.

\noindent Key Words:
quasars: absorption lines $-$ diffuse radiation \\

\section{Introduction}

The shape and the intensity of the intergalactic UV background radiation are
crucial factors in determining the ionization balance of the intergalactic
medium and therefore influence the structure formation in the universe.
Knowledge of this radiation field is thus necessary for the understanding of the
early universe. AGNs are believed to be the major contributors to this
background, though a significant contribution from star forming galaxies can not
be ruled out. Detailed calculations of the propagation of AGN like ionizing
radiation through intergalactic space, taking into account the absorption and
reradiation by the galactic and intergalactic material, have been carried out by
Haardt and Madau (1996, hereafter HM96).  They have determined the frequency and
redshift dependence of the background.  Observationally, the intensity of the
radiation has been determined in recent years, by studying the proximity effect
in the Lyman alpha forest of the absorption lines in the spectra of QSOs
(Bajtlik, Duncan \& Ostriker 1988).  This analysis is insensitive to the shape
of the radiation (Bechtold, 1994 and Das and Khare, 1997).  Values of the
intensity of the background at the Lyman limit, J$_{\rm \nu_{LL}}$, obtained by
Bechtold (1994) and Cooke et al (1996) are considerably higher than the value
expected from the distribution of visible QSOs.  Several sources of uncertainty
in the value of the flux obtained by the proximity effect analysis have been
considered by various authors (Bechtold, 1994, 1995, Srianand and Khare, 1995,
Das and Khare, 1997).  It has also been suggested (Fall \& Pei 1993) that the
actual number of QSOs may be larger than their observed number and that several
QSOs may be rendered invisible due to dust extinction in the intervening
absorbers.  It is possible that the background radiation gets a significant
contribution from star forming galaxies (Madau \& Shull 1996, Giroux \& Shapiro
1996, Khare and Ikeuchi, 1998).  Here we try to obtain an independent estimate
of the background flux by studying the ionization state of the QSO absorption
systems for which an estimate of the particle density is available from the
observations of the fine structure excited lines of C II.  Where ever possible,
we also try to estimate the contribution of the galactic flux to the total
ionizing flux for the systems.  In section II we present our analysis, the
results are discussed in section III.

\section{Data and Analysis}

Several absorption systems reported in the literature, have the absorption lines
of C II, C II*, C IV, H I, Si II, Si IV etc.  Particle density in the absorber
can be obtained from the column densities of C II and C II*.  The column density
ratio of C II to C IV is a good indicator of the ionization parameter as it
changes rapidly with change in the ionization parameter, $\Gamma= {\Phi\over{\rm
c\;n_H}}$, where $\Phi$ is the flux of the ionizing radiation i.e.  the number
of photons cm$^{-2}$ s$^{-1}$, $\rm n_H$ is the particle density and c is the velocity of light.  This ratio is
insensitive to the particle density and the abundance of carbon, but is very
sensitive to the neutral hydrogen column density, N$_{\rm H\;I}$, in the
absorber, specially for N$_{\rm H\;I}> 10^{17}$ cm$^{-2}$ (Bergeron and 
Stasinska
1986 ).  This particular column density ratio is also very sensitive to the
shape of the ionizing radiation.  This is shown in Fig.1 which shows the ratios
for neutral hydrogen column densities of 10$^{17}$ cm$^{-2}$ and 10$^{19}$
cm$^{-2}$ for (i) spectral shape as given by HM96 for a redshift of 2.5, (ii)
galactic spectra as taken from Bruzal (1983) and (III) a combination of both the
above with the AGN flux taken to be same as the actual value given by HM96 for
z=2.5.  These results have been obtained from the code 'CLOUDY', kindly supplied
to us by Prof.  Ferland.  The galactic spectra produces much smaller values of
the ratio compared to the AGN or power law spectra.  This is due to the fact
that the galactic spectra has much smaller number of photons having sufficient
energy to produce C IV.  Thus it is necessary to know the shape of the ionizing
radiation to determine the ionization parameter from the C II to C IV ratio.  In
order to get information about the intensity and the shape of the radiation field we can make use of the column density
ratio of other ions.  Si II to Si IV ratio is an useful ratio for this purpose.
In Fig.  2 we have plotted the ratio of column densities of Si II to Si IV for
the three cases listed above.  This ratio is not very sensitive to the shape and
can be used with the C II to C IV ratio to constrain the shape as well as the
intensity of the ionizing radiation.  Fe II to Fe III and Al II to Al III ratios
can also be used for this purpose.  These ratios are also plotted in Fig.  2.
We have, therefore, selected from the literature, absorption systems for which
the column densities of H I, C II and C II*, or limits on their values, have 
been
determined and for which additional ions like the C IV, Si II, Si IV, Fe II, Fe
III etc are also available.  The details are given in Table 1.

The fine structure excited level of C II is primarily populated by collisions
with electrons for the absorption systems considered here, which are either the
Lyman limit or the damped Lyman alpha systems, while the collisions with H I and
also the collisional deexcitations by electrons can be ignored (Bahcall and
Wolfe, 1968, Morris et al 1986).  As we are considering the absorption systems
at high redshifts, the excitation of the finestructure level by the absorption
of the cosmic microwave background photons may be important.  In order to check
this we calculated the excitation temperatures of C II* from the observed column densities.  These are much higher
than the corresponding temperatures of the microwave background at the redshifts
of the absorbers as can be seen from Table 2.  We have therefore, ignored the
excitation by the microwave background photons and have assumed the electron
densities (assumed to be equal to the proton density) to be given by n$_{\rm e}$
= 21 [${\rm N_{\rm C\;II*}}\over{\rm {N_{C\;II}}}$] (Morris et al 1986).  The
hydrogen density obtained by adding the neutral hydrogen density to the electron
density, corrected for the electrons coming from He II and He III, for each
system is also given in Table 2.  For a few of the systems, only an upper limit
on C II*/C II was available.  For these systems only an upper limit on the
particle density could be obtained.  For these systems we have assumed a lower
limit on particle density to be 0.045 cm$^{-2}$, which is smaller than all the
lower limits to the particle densities obtained for the systems considered here
and is also considerably lower than the mean interstellar particle density.  For
each absorption system we have constructed a number of photoionization models
for the observed neutral hydrogen column density and different spectral shapes.
The AGN spectral shape at the redshift of the absorption system is taken from
HM96.  For some of the systems only the total neutral hydrogen column density is
available, while C II* has been observed in some particular velocity component of the absorption line.
For such systems we have assumed the neutral hydrogen column density in
individual components to be in the same ratio as the Si II column
densities, as the ionization potential of Si II is close to that of H I.
Details of the models and comparison of their predictions with the observations
are discussed below for individual absorption systems.  The results are given in
Table 2.  Note that all the absorption systems are sufficiently far away from
the respective QSOs (relative velocity is greater than 15000 km s$^{-1}$) and
can be considered to be intervening (however, see Richards et al, 1999) so that
the radiation of the parent QSO can be ignored.  In the following analysis we
have only used the ratios of the column densities of different ions of the same
element.  Our conclusions are, therefore, independent of the assumed values of
chemical abundances.

\subsection{z=1.7765 system towards Q1331+170} This QSO has been observed by
Kulkarni et al (1996).  However, for this system C IV column density has not
been reported and so we could not constrain the spectral shape.  Errors on
column densities have not been reported by the authors and we assumed errors of
25$\%$ in the column densities.  Neutral hydrogen column density in the
component showing C II* has been obtained from the total H I column density
(Green et al 1995) by assuming the H I column densities to be in the same ratio
as the Si II column densities.  Analysis of the Si lines assuming HM96 spectral
shape for z=2 yields -2.4 $> \Gamma > -2.6$ giving 3.2 $\times 10^7 \le \Phi \le
1.4\times 10^8$.  This is considerably higher than the value of HM96 flux at the
redshift of the absorption system. It is, however, possible that the excess flux comes
from the galaxies.  We explored this possibility by constructing photoionization
models with the shape as well as the intensity of the background as given by
HM96 at the redshift of the absorber, the rest coming from the galaxies.  For
these models the limits become -2.7 $> \Gamma > -2.9$ giving 1.5 $\times 10^7
\le \Phi_{\rm G} \le 6.9 \times 10^7$, $\Phi_{\rm G}$ being the galactic flux in
cm$^{-2}$ s$^{-1}$.

\subsection{z=2.279 system towards Q2348-14} This QSO has been observed by
Pettini et al (1994).  An upper limit on the column density ratio of Si II to Si
IV gives, for HM96 spectral shape for z=2.5, a lower limit of -2.2 for log
$\Gamma$.  However, as only an upper limit on the particle density could be
obtained, this can not be converted to a limit on the flux.  Assuming the
minimum value of the particle density to be 0.045 cm$^{-3}$, which is smaller
than the observed upper limit by 1.2 dex, requires the flux to be larger than
8.5 $\times 10^6$.  The column density ratio of Al II to Al III gives -1.3 $>$
log $\Gamma > -1.4$, giving $ 5.3\times 10^7 < \Phi < 1.2 \times 10^9$.  Taking
the actual value of the HM96 flux and assuming the rest of the contribution from
the galactic sources requires -1.4 $>$ log $\Gamma > -1.5$, giving the galactic
flux to be between 4.1 $\times 10^7$ and 9.7 $\times 10^8$.

\subsection{z=2.638 system towards PKS 2126-158}

This system has been observed by Giallongo et al (1993) and has 7 components
spread over a velocity width of 270 km s$^{-1}$.  A neutral hydrogen column
density for the whole system has been obtained by Young et al (1979) to be 1.1
$\times 10^{19}$ cm$^{-2}$.  We have assumed the neutral hydrogen column density
in individual components to be in the same ratio as the Si II column densities.
C II* is observed in two of the components.  These are considered below.

\noindent (i) z=2.6376:  We obtain log N$_{\rm H I}$ = 17.92.  As Si IV lines
are not observed we take a 3 $\sigma$ upper limit on equivalent width, which
translates to an upper limit of 10$^{13.23}$ cm$^{-2}$ on the column density of
Si IV, assuming a velocity dispersion parameter of 24.2 km s$^{-1}$, same as
that for C IV.  Assuming AGN shape, we get -3.1 $>$ log $\Gamma$ $>$ -3.3,
giving 1.6 $\times 10^6 \le \Phi \le 1.4 \times 10^8$.  Taking the AGN flux to
be that given by HM96 for z=2.5, and additional flux from the galaxies, the 
observed column density ratios can not be explained and a
minimum flux of 9.4 $\times 10 ^5$ from the AGN is required, needing a total of
-2.9 $>$ log $\Gamma$ $>$ -3.1, giving 1.6 $\times 10^6 \le \Phi_{\rm G} \le 2.3
\times 10^8$.  Note that the Al II to Al III ratio can not, however be explained
by the same range of $\Gamma$ for any shape and requires log $\Gamma \le -3.3$.

\noindent (ii) z=2.6364.  We obtain log N$_{\rm H I}$ = 16.4.  As Si II and Si
IV lines have not been detected for this system, we could not obtain any
constraints on the relative contribution from galaxy to the radiation flux.
Taking all of the radiation to be AGN type, we obtained limits on the ionization
parameter to be -2.7 $\ge$ log $\Gamma \ge -3.3$ so that the flux lies between
1.05$\times 10^6$ and 6.97$ \times 10^8$.  Taking the actual value of the AGN
flux (HM96 value at z=2.5), the rest coming from galaxy, requires log $\Gamma
\ge -1.0$, giving $\Phi_{\rm G} \ge 1.2 \times 10^{10}$.

\subsection{z=2.6522 system towards Q2231-00}

This Lyman limit system has been analysed in details by Prochaska (1999).  We
have reanalysed this system taking the shape of the background to be that given
by HM96 for z=2.5.  We determined the column density ratios for various ions for
log N$_{\rm H I}$ of 19.12.  The Fe II to Fe III and Si II to Si IV column
density ratios constrain the ionization parameter to between -2.4 to -2.55
resulting in a background flux between 3.9$\times 10^6$ and 8.5$\times 10^6$.
This is considerably higher than the value of 8.59$\times 10 ^5$ for HM96 for
z=2.5.  Models with the shape as well as the intensity of the background as
given by HM96 for z=2.5, the rest of the flux coming from galaxy fail to yield a
result as the column density ratios of Si and Fe can not simultaneously be
produced by a single value (range) of ionization parameter.  By gradually
increasing the value of the flux contributed by the AGN background above the 
HM96 value, we find that
a minimum AGN flux of 3.3 $\times 10^{6}$ was needed to explain the ion ratios,
for which we get -2.4 $>$ log $\Gamma$ $>$ -2.6, requiring $\Phi_{\rm G}$ to be
between 6.7 $\times 10^{5}$and 5.2 $\times 10^{6}$.  Thus the minimum of AGN
type flux required is more than a factor of 3.8 larger than that obtained by
HM96.

\subsection{z=2.844 system towards HS1946+7658} This system has been observed
 and analysed by Fan and Tytler (1994).  Cloudy models with HM96 spectral shape
 at z=3 give -2.3 $>$ log $\Gamma >$ -2.8, giving $10^8 \le \Phi \le 10^9$.  A
 minimum flux of 8.4$ \times 10^7$ of the HM96 type is needed, with -1.9 $\ge$
 log $\Gamma \ge -2.6$, giving 1.9 $\times 10^8 \le \Phi_{\rm G} \le 3.9 \times
 10^9$.  Thus almost all of the flux is being contributed by galaxy.  Note that
 the flux for z=3 of HM96 is 7.8$\times 10^5$.

\subsection{z=3.025 system towards Q0347-38} This QSO has been observed by
Prochaska and Wolfe (1999).  Only a lower limit is available for the C II to C
IV ratio, so that the galactic fraction of the flux could not be constrained.
AGN shape for z=3.0 for the radiation gives -2.6 $>$ log $\Gamma > -2.8$, giving
2.1 $\times 10^6 \le \Phi \le 9.8 \times 10^7$, for the assumed minimum value of
the particle density.  Taking the actual value of the HM96 flux for z=3 and
assuming the rest of the contribution from the galactic sources requires -2.5
$\ge$ log $\Gamma \ge -2.7$, giving 1.9 $\times 10^6 \le \Phi_{\rm G} \le
1.2\times 10^8$, for the assumed minimum value of n$_{\rm H}$. 

\subsection{z=3.6617 system towards Q2212-1626} This QSO has been observed by Lu
et al (1996).  Only a lower limit is available for the C II to C IV ratio, so
that the galactic fraction of the flux could not be constrained.  AGN shape for
z=3.5 for the radiation gives -2.5 $>$ log $\Gamma > -2.6$, giving $ 3.4\times
10^6 \le \Phi \le 1.1 \times 10^8$ for the assumed minimum value of n$_{\rm H}$.
Taking the actual value of the HM96 flux at z=3.5 and assuming the rest of the
contribution from the galactic sources requires -2.4 $>$ log $\Gamma > -2.5$,
giving $3.8 \times 10^6 \le \Phi_{\rm G} \le 1.4\times 10^8$, for the assumed
minimum value of n$_{\rm H}$.  Note that the value of flux for HM96 for z=3.5 is
5.0 $\times 10^5$.

\subsection{z=4.0803 system towards Q2237-0608} This QSO has been observed by Lu
et al (1996). Only a lower limit is available for the C II to C IV ratio, so
that the galactic fraction of the flux could not be constrained.  AGN shape for
z=4 for the radiation gives -2.6 $>$ log $\Gamma > -2.7$, giving $2.6 \times
10^6 \le \Phi \le 2.1 \times 10^7$ for the assumed minimum value of n$_{\rm H}$.
Taking the actual value of the HM96 flux at z=4 and assuming the rest of the
contribution from the galactic sources requires -2.4 $\ge$ log $\Gamma > -2.5$,
giving $4.0\times 10^6 \le \Phi_{\rm G} \le 3.3\times 10^7$ , for the assumed
minimum value of n$_{\rm H}$.  Note that the value of flux for HM96 for z=4 is 2
$\times 10^5$

\section{Discussion}

For 5 of the systems we could derive the range of flux values assuming the
radiation to be AGN type.  All of these are higher than the corresponding HM96
values by minimum factors ranging from 1.2 to 158.  Note that we have taken into
account the uncertainties in the column densities of all the ions which is the
reason for obtaining large ranges for the flux values.  For three of these
systems we could obtain a minimum value for the flux of the AGN background.
These values are 1.1, 3.8 and 98.8 times higher than the HM96 values at the
appropriate redshifts.  For these systems a large flux is needed from galaxies.
For 4 other systems a lower limit to the flux could only be obtained with an
assumption of the lower limit on the particle density to be 0.045 cm$^{-3}$, which is
about half of the mean interstellar value of the particle density and which
indicates that the actual C II column densities are higher than the observed
lower limits by 0.64 to 1.37 dex.  This is a reasonable lower limit as the
systems being considered are Lyman limit or damped Lyman alpha systems and also
as this value is considerably lower than the range of density values for systems
for which the values could be obtained from the observations.  For these systems,
the required values of flux are higher than the HM96 values by minimum factors
of 2.7 to 62.  On the other hand, assuming the AGN flux to be that given by
HM96, and assuming the rest of the required flux to be of local, galactic
origin, very high galactic flux is required.  For most of the systems, this high
flux requires the absorption systems to be present with in 100 parsecs of
typical O stars.  The typical radius of the Stromgren spheres of these stars is
of the same order, indicating that the absorption systems are inside the H II
regions.  Such conclusions have earlier been rejected on the basis of
statistical arguments about the properties of the absorption systems (Srianand
and Khare, 1994).  The flux could come from QSOs which happen to lie close to
the lines of sight at redshifts similar to the redshifts of the absorption
systems.  We have searched the catalogues for presence of any such QSOs near the
line of sight to Q2231-00.  However, no QSO is found to lie closer than 1000 Mpc
to the line of sight within the required redshift range.  The high values of the
flux indicated by our analysis for almost all the systems, may be interpreted to
indicate the presence of an unseen population of dust extinct QSOs.

Note that in all our analysis we have assumed that all the ions producing
absorption in a given velocity range in an absorption system are physically
located in the same region (cloud).  This may not be always valid.  Kirkman and
Tytler (1999) and Churchill and Charlton (1999) have found
evidence for ions with the same velocity structure in their absorption lines
belonging to a given redshift system, arising in physically different gaseous
components.  If C IV ions are from a more widely distributed component, then,
the C II/C IV column density ratio in the region of interest will be smaller and
may require lower values of the flux.

\section*{acknowledgement} This work was partially supported by a grant (No.
SP/S2/013/93) by the Department of Science and Technology, Government of India.
A.S. is supported by a C.S.I.R. fellowship.

\noindent{\bf Figure Captions\\}
\noindent {\bf Figure 1.}
Column density ratios of C II to C IV as a function of the ionization parameter, 
for different shapes of the background radiation spectrum as explained in the 
text. The solid lines are for N$_{\rm H I}$=10$^{17}$ cm$^{-2}$ and dashed lines 
are for  N$_{\rm H I}$=10$^{19}$ cm$^{-2}$.\\
\noindent {\bf Figure 2.} 
Column density ratios of Si II to Si IV, Al II to Al III and Fe II to Fe III. 
for the three shapes of the background radiation spectrum as explained in the 
text, for N$_{\rm H I}$=10$^{18}$ cm$^{-2}$. Solid, dotted and dashed lines are 
for galactic spectra, AGN (HM96) spectra and AGN together with the galactic 
spectra respectively.\\
\end{document}